\documentclass[11pt]{article}
\usepackage{amssymb}
\usepackage{epsfig}

\title{Computing with spins: From classical to quantum computing}

\author{S. Bandyopadhyay \\
\\Department of Electrical Engineering and Department of Physics \\ 
Virginia Commonwealth 
University, Richmond, VA 23284, USA}

\begin{document}
\maketitle

\begin{abstract}

This article reviews the use of single electron spins to compute. In classical 
computing schemes, a binary bit is represented by the spin polarization of a 
single electron confined in a quantum dot. If a weak magnetic field is present, 
the spin orientation becomes a binary variable which can encode logic 0 and 
logic 1. Coherent superposition of these two polarizations  represent a qubit. 
By engineering the exchange interaction between closely spaced spins in 
neighboring quantum dots, it is possible to implement either classical or 
quantum logic gates.

\end{abstract}

\pagebreak

\section{Introduction}

The visionary  who first thought of using the spin polarization of a single 
electron 
 to encode a binary bit of information has never been 
identified conclusively. Folklore has it that Feynman mentioned this 
notion in casual conversations (circa 1985), but to this author's knowledge
there did not exist concrete schemes for implementing spintronic logic gates 
till the mid 1990s. Encoding information in spin may have certain advantages. 
First, there is the possibility of lower power dissipation in switching logic 
gates. In charge based devices, such as metal oxide semiconductor field effect 
transistors, switching between logic 0 and logic 1 is accomplished by moving 
charges into and out of the transistor channel. Motion of charges is induced by 
creating a potential gradient (or electric field). The associated potential 
energy is ultimately dissipated as heat and irretrievably lost. In the case of 
spin, we do not 
have to {\it move} charges. In order  to switch a bit from 0 to 1, or vice 
versa, we merely have to toggle the spin. This may require much less energy. 
Second, spin does not couple easily to stray electric fields (unless there is 
strong spin-orbit interaction in the host material). Therefore, spin is likely
to be relatively immune to noise. Finally, it is possible that spin devices may 
be 
faster. If we do not have to move electrons around, we will not be limited by 
the transit time of charges. Instead, we will be limited by the spin flip time, 
which could be smaller. 

\section{Spintronic classical (irreversible) logic}  

In 1994, we proposed a concrete scheme for realizing a classical universal logic 
gate (NAND) using three spins placed in a weak magnetic field \cite{bandy}.
By ``three spins'', we mean the spin orientations of three conduction band 
electrons, each confined in a semiconductor quantum dot. The system is shown 
schematically in Fig. 1. Exchange interaction is allowed only between nearest 
neighbor spins (second nearest neighbor interaction is considered too weak to 
have any
effect). Because of the magnetic field, the spin orientation in any quantum dot 
becomes a {\it binary variable}. The spin polarization is either along the 
magnetic field, or opposite to the field. To understand this, consider the 
Hamiltonian of an isolated dot:
\begin{equation}
H = H_0 - (g/2)\mu_B {\bf \sigma} \cdot {\bf B}
\end{equation}
where $H_0$ is the unperturbed Hamiltonian in the absence of the magnetic field, 
${\bf B}$ is the magnetic field, $g$ is the Land\'e g-factor of the quantum dot 
material, $\mu_B$ is the Bohr magneton, and ${\bf \sigma}$ is the Pauli spin 
matrix. If the magnetic field is directed along the z-direction, then
\begin{equation}
H = H_0 - (g/2) \mu_B \sigma_z B
\end{equation}
Diagonalizing the above Hamiltonian yields the eigenspinors (1,0) and (0,1) 
which are +z-- and -z--polarized spins. Therefore, the spin orientation is a 
binary 
variable; it is either parallel or anti-parallel to the applied magnetic field.

\begin{figure}[h]
\centerline{\psfig{figure=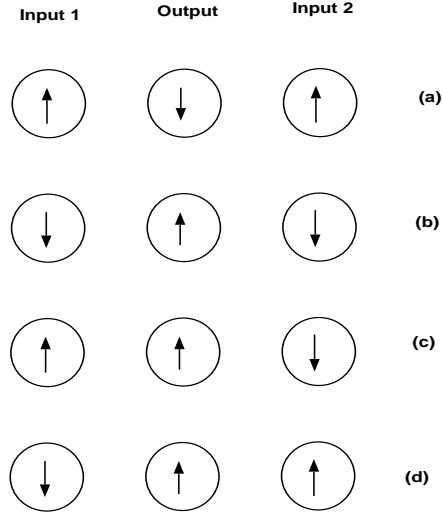,height=3.4in,width=3.4in}}
\caption{Spintronic realization of a {\it single} NAND gate. The spin 
configurations when (a) both inputs are logic 1 (upspin), (b) both inputs are 
logic 0 (downspin), (c) and (d) one input is logic 1 and the other is logic 0.}
\end{figure}

In the presence of exchange interaction between two electrons confined to 
two {\it separate} potentials (such as two different quantum dots), the 
anti-ferromagnetic ordering, or the singlet state, (i.e. two neighboring spins 
are 
anti-parallel) is preferred over the ferromagnetic ordering, or triplet state  
(two spins are parallel) \cite{ashcroft}. We will assume that the tendency to 
preserve this anti-ferromagnetic ordering is {\it stronger} than the tendency 
for all spins to line up along the magnetic field. This merely requires that the 
exchange splitting energy $J$ (energy difference between triplet and singlet 
states) exceed the Zeeman splitting energy $g \mu_B B$. We ensure this by 
reducing the potential barrier between neighboring dots to enhance the exchange, 
while at the same time, making the magnetic field sufficiently weak to reduce 
the Zeeman energy.
Under this scenario, the ground state of the 
array has the spin configuration shown in Fig. 1(a). We will call ``upspin'' the 
spin orientation directed along the magnetic field and ``downspin'' the opposite 
orientation.

We encode logic 1 in the upspin state. Furthermore, we will consider the two 
edge dots in Fig. 1(a) as input ports to a logic gate, and the middle dot as the 
output port.
It is obvious that when the two inputs are logic 1,   the output will be logic 0 
when the system reaches ground state (anti-ferromagnetic ordering).

Next, consider the situation when the two inputs are logic 0 (see Fig. 1(b)). 
The output must be 
logic 1 in order to conform to the anti-ferromagnetic ordering. However, there 
is a subtle issue. Fig. 1(b) is 
actually {\it not} the ground state of the system, Fig. 1(a) is. This is because 
of the weak magnetic 
field. The difference between Fig. 1(a) and Fig. 1(b) is that in the former 
case, {\it two} spins are aligned parallel to the magnetic field, while in the 
latter, {\it two} spins are aligned anti-parallel to the magnetic field. 
Therefore, if 
the system is left in the state of Fig. 1(b), it must ultimately decay to the 
state in Fig. 1(a), according to the laws of thermodynamics. But that may take a 
very long time because of three reasons. First, the system must emit some energy 
carrying entity to decay. This entity is most likely a phonon. However, phonon 
emissions in quantum dots are suppressed by the ``phonon bottleneck'' effect 
\cite{benisty}. Second, phonons do not couple easily to spin unless we have a 
strongly pyroelectric material as the host. Finally, if spins flip one at a time 
(all three spins flipping simultaneously is very unlikely), then in order to 
access the state in Fig 1(a), the state in Fig. 1(b) will have to go through a 
state where two neighboring spins will be parallel. Such a state is much higher 
in energy than either Fig. 1(a) or Fig. 1(b). Therefore, Fig. 1(a) and Fig. 1(b) 
are separated by an energy barrier, making Fig. 1(b) a long lived metastable 
state. As long as the input bit rate is high enough so that inputs change much 
more rapidly than the time it takes for the metastable state to decay to the 
global ground state of Fig. 1(a), we need not worry about this issue.

What happens if one of the inputs is logic 1, and the other is logic 0 as shown 
in Fig. 1(c)? Here the magnetic field comes in handy to break the tie. In this 
case, logic 1 is preferred as the output since the all other things being 
equal, a spin would prefer to line up parallel to the magnetic field, rather 
than anti-parallel. Thus, when either input is logic 0, the ouput is logic 1.
We have realized the truth table in Table 1.

\begin{table}
\begin{center}
\caption{Truth table of a spintronic NAND gate}
\begin{tabular}{|c|c|c|}
\hline
Input 1 & Input 2 & Output \\
\hline
1 & 1 & 0 \\
1 & 0 & 1 \\
0 & 1 & 1 \\
0 & 0 & 1 \\
\hline
\end{tabular}
\end{center}
\end{table}

The reader will recognize that this is the truth table of a NAND gate, which is 
one of two universal Boolean logic gates. Since we can realize a NAND, we can 
realize any arbitrary Boolean logic circuit (combinational or sequential) by 
connecting NAND gates. A number of different logic devices (half adders, 
flip-flops, etc.) were designed and illustrated in ref. \cite{bandy}.

These devices have been extensively studied by others \cite{molotkov,bychkov}
using time independent simulations. The time-independent 
simulations address the steady state behaviors and therefore do not directly 
reveal a serious problem with these devices that was already recognized in ref. 
\cite{bandy}. In the following section, we explain this problem. 

\section{Problem: Lack of unidirectionality}

At the time these logic gates were proposed, it was also realized that they have 
a severe shortcoming that precludes their use in pipelined architectures 
\cite{bandy}. To understand the nature of the problem, consider three inverters 
(NOT gates) in series. A single NOT gate is the simplest device; it is realized 
by two exchange coupled spins, one of which is the input and the other is the 
output. Because of the anti-ferromagnetic ordering, the output is always the 
logic complement of the input.

\begin{figure}[h]
\centerline{\psfig{figure=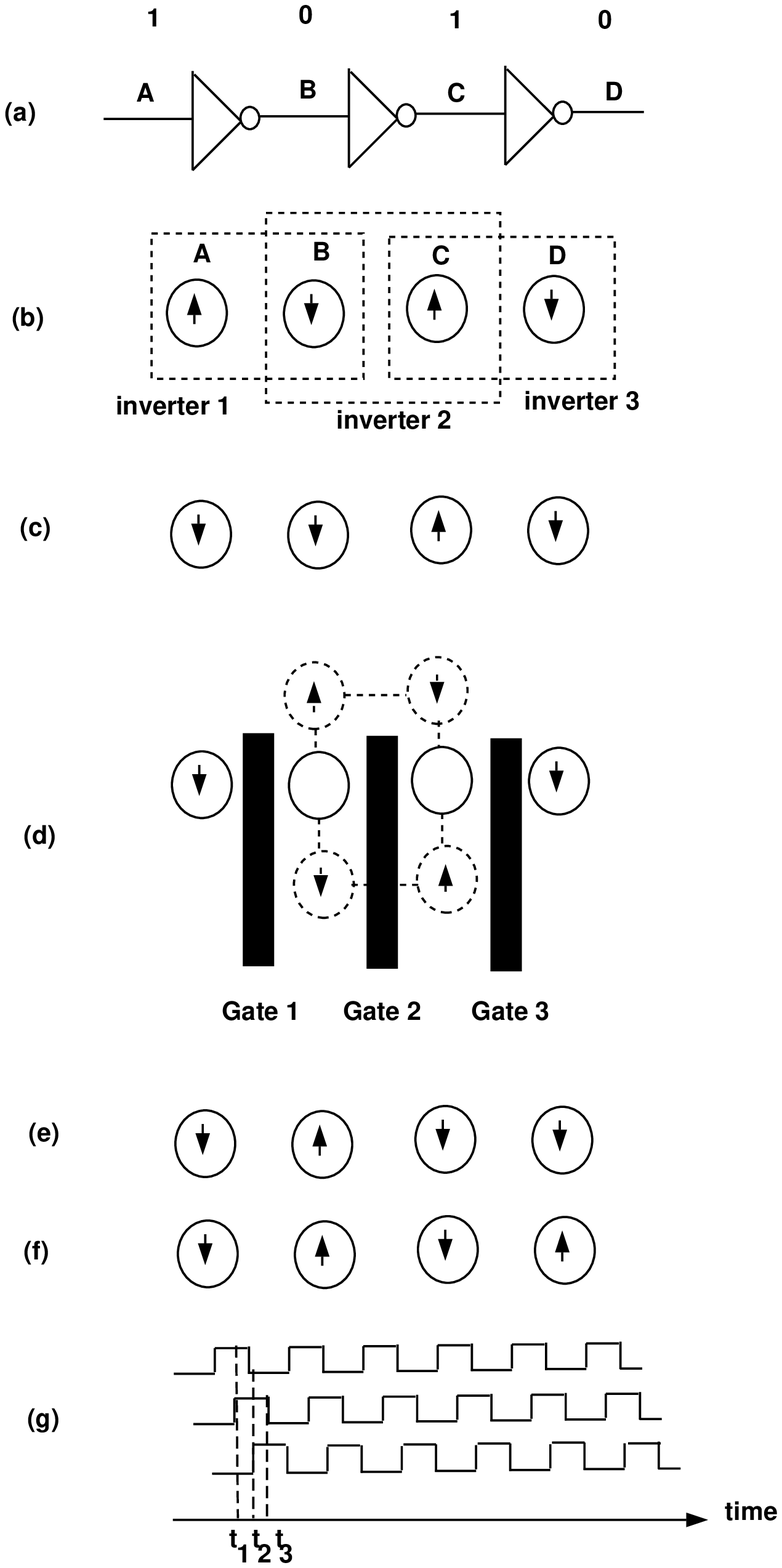,height=7.2in,width=4.8in}}
\caption{(a) Three conventional inverters in series with the logic states at 
four 
different nodes (A, B, C and D) shown. (b) Spintronic realization of the circuit 
in Fig. 2(a), (c) The spin states at time $t$ = 0+, immediately after the 
leftmost spin is flipped by an external input, (d) Configurations attainable 
with single phase clocking. There is an ambiguity as shown in the two branches 
(broken lines). Configurations attainable with multi-phase clocking: (e) when 
the first two gates are raised in potential, (f) when the next two gates are 
raised in potential, (g) the three clock pulse 
trains required in this case are shown. The first gate is tied to the first 
train, the second to the second train and the third to the third train. Then the 
fourth gate is again tied to the first train and the pattern repeated. The 
period between $t_1$ and $t_2$ corresponds to Fig. 2(e) and the period between 
$t_2$ and $t_3$ corresponds to Fig. 2(f).}
\end{figure}

Fig. 2(a) shows three conventional inverters in series and Fig. 2(b) shows the 
corresponding spintronic realization. The input to the first inverter is logic 1 
and the 
output of the last inverter is  logic 0, as it should be. But now, let us 
suddenly change the input at the first inverter to logic 0 at time $t$ = 0.
The situation at time $t$ = 0+ is shown in Fig. 2(c). We expect that ultimately
the output of the last inverter will become logic 1. Unfortunately, this cannot 
happen. In Fig. 2(c), the second spin from the left finds its left neighbor 
asking it to flip (because of the exchange interaction that enforces 
anti-ferromagnetic ordering between two neighboring spins) while its right 
neighbor is asking it to stay put because of the same exchange interaction. Both 
influences -- from the left and from the right -- are exactly equally strong and 
the second cell is stuck in a logic indeterminate state that it cannot get out 
of \cite{bandyo}. Rolf Landauer later termed it a metastable state that 
prevents decay to the desired ground state \cite{landauer}. In fact, if we take 
the external magnetic field into account, then there is a preference for the 
second cell to actually {\it not} flip in response to the input since there is a 
slight preference for the upspin state because of the external magnetic field. 
In this case, the logic signal cannot propagate from the input to the output and 
the circuit simply does not work! Similar situations were examined in ref. 
\cite{lusth}.
The real problem 
is that exchange interaction is {\it bidirectional} which cannot ensure 
{\it unidirectional} flow of logic signal from the input to the output of the 
logic device. This unidirectionality is a required attribute of any logic 
device (for the five necessary requirements of a classical logic device, see 
ref. \cite{hodges}). We can think of desperate measures to enforce the 
unidirectionality. For example, we can claim that if we hold the input at the 
first inverter (leftmost cell in Fig. 2(c)) to logic 0, and do not let go, then
the second cell which is equally likely to follow its left neighbor and right 
neighbor, will have no option but to ultimately follow its left neighbor since 
it is adamant and persistent (we are not letting go of the input). This will 
happen in spite of the magnetic field since the exchange energy is larger than 
the Zeeman splitting. In this case, we are trying 
to enforce unidirectionality via the input signal itself (note that the input 
device does indeed break the inversion symmetry of the system in Fig. 2(c)). 
This possible remedy was studied theoretically in ref. \cite{anant} which 
reached the conclusion that it does not always work. In fact, the process of 
logic signal propagation under this scenario is inefficient thermally assisted 
random walk and the final logic state, if reached, can be destroyed by thermal 
fluctuations.
The idea of using the input device to enforce unidirectionality was also 
implicitly used in the experiment of ref. \cite{cowburn}. While this may work 
for a few cells (as it did in ref. \cite{cowburn}), it will obviously not work 
for a large number of cells since the influence of the input decays with 
increasing distance from the input. Ultimately, the remote cells that are 
far from the input, will not feel the input's effect and remain stuck in 
metastable states, producing wrong answers to simple logic problems.

\subsection{Possible solution}

In ref. \cite{bandy}, one solution that was offered to break this impasse was to 
progressively 
increase the distance between cells. This makes the influence of the left 
neighbor always stronger than that of the right neighbor since the strength 
of the exchange interaction has an exponential dependence on the separation
between neighboring cells. This is not an elegant solution since ultimately 
the exchange splitting energy will become smaller than the Zeeman splitting 
energy, at which point the paradigm will no longer work.

In 1996, we proposed a more elegant solution \cite{jjap}. This was inspired by 
the 
realization that in charge coupled device (CCD) arrays, there is no inherent 
unidirectionality, yet charge is made to propagate from one device to the next 
unidirectionally. This is achieved by {\it clocking}. We mentioned that 
unidirectionality can be imposed in time or space \cite{jjap} and clocking 
imposes unidirectionality in time. However, a cursory examination revealed
that normal clocking will not work in our case. Say, we put gate pads on the 
barriers between neighboring cells. Initially, all the the barriers are
high and opaque so that there is no overlap between the wavefunctions of 
adjacent electrons and hence no exchange interaction between neighboring spins.
Now, we lower the first barrier by applying a positive potential to gate 1 as 
shown 
in Fig. 2(d). This allows the wavefunctions of electrons on either side of the 
gate pad to leak out into the barrier, overlap. and cause an exchange 
interaction. Exchange  causes the second spin to assume a polarization 
anti-parallel to the input spin orientation since the singlet state is lower in 
energy than the triplet. In other words, the second cell switches. At this 
point, if we let go of the input, raise the first barrier back up, and lower the 
second barrier by applying a positive potential to gate 2, then either the third 
cell switches in response to the second (which is shown
in the upper branch), or the second cell switches in response to the third
(which is shown in the lower branch). The upper branch is the 
desired state, but it is equally likely that the lower branch will result 
since both branches obey the ant-ferromagnetic ordering between the two exchange 
coupled cells (cell 2 and 3). Therefore, a simple sequential clock will not 
work. What is 
required is that both gate 1 and gate 2 have positive potentials while the 
input is applied. Now the first three cells assume the correct polarizations as 
shown in Fig. 2(e).
Then, the input is removed, gate 1 is returned to zero potential and positive 
potentials are
applied to gates 2 and 3. This causes the first four cells to assume the correct 
polarization, and so on. This situation is shown in Fig. 2(f) and is the desired 
configuration. Thus by lowering {\it two adjacent barriers} pairwise 
at the same time, we can propagate the input state through a linear array. In 
other words, we will need a {\it three-phase clock}, a single phase will 
not work. The three clock pulse trains for a three-phase clock are shown in Fig. 
2(g). Each train is phase shifted from the previous one by $\pi/2$ radians. Such 
a situation is not unusual since charge coupled device 
arrays also need a multi-phase clock (push clock, drop clock) to work 
\cite{yaeger}.

While, multi-phase clocking can make these devices work, it is hardly an 
attractive solution since one needs to fabricate gate pads between every two 
cells. The separation between the cells may need to be $\sim$ 5 nm
in order to have sufficient exchange coupling. Aligning a gate pad to within a 
space of 5 nm is a major lithography challenge. Furthermore, the gate potentials 
are 
lowered and raised by moving charges into and out of the gate pads, leading to 
considerable energy dissipation that completely negates the advantage of using 
spins. Therefore, these devices 
present interesting physics, but at this time, do not appear to be serious 
candidates for practical applications. 

\section{Using spin as a qubit}

So far, we have discussed the use of spin in classical irreversible logic gates. 
These logic gates dissipate a minimum of $kTln2$ amount of energy per bit flip 
\cite{landauer2}.
Let us assume that we can make quantum dots with a density of 10$^{12}$ 
cm$^{-2}$. Quantum dots self assembled by electrochemical techniques in our own 
lab (and in many other labs) can achieve this density today. We show a raw 
atomic force micrograph of quantum dots self assembled in our lab in Fig. 3. The 
dark areas are the dots and the surrounding light areas are the barriers. The 
dot diameter in this micrograph is 50 nm and the dot density is close to 
10$^{11}$ cm$^{-2}$. By using slightly different synthesis conditions, we can 
actually achieve densities exceeding 10$^{12}$ cm$^{-2}$.

\begin{figure}[h]
\centerline{\psfig{figure=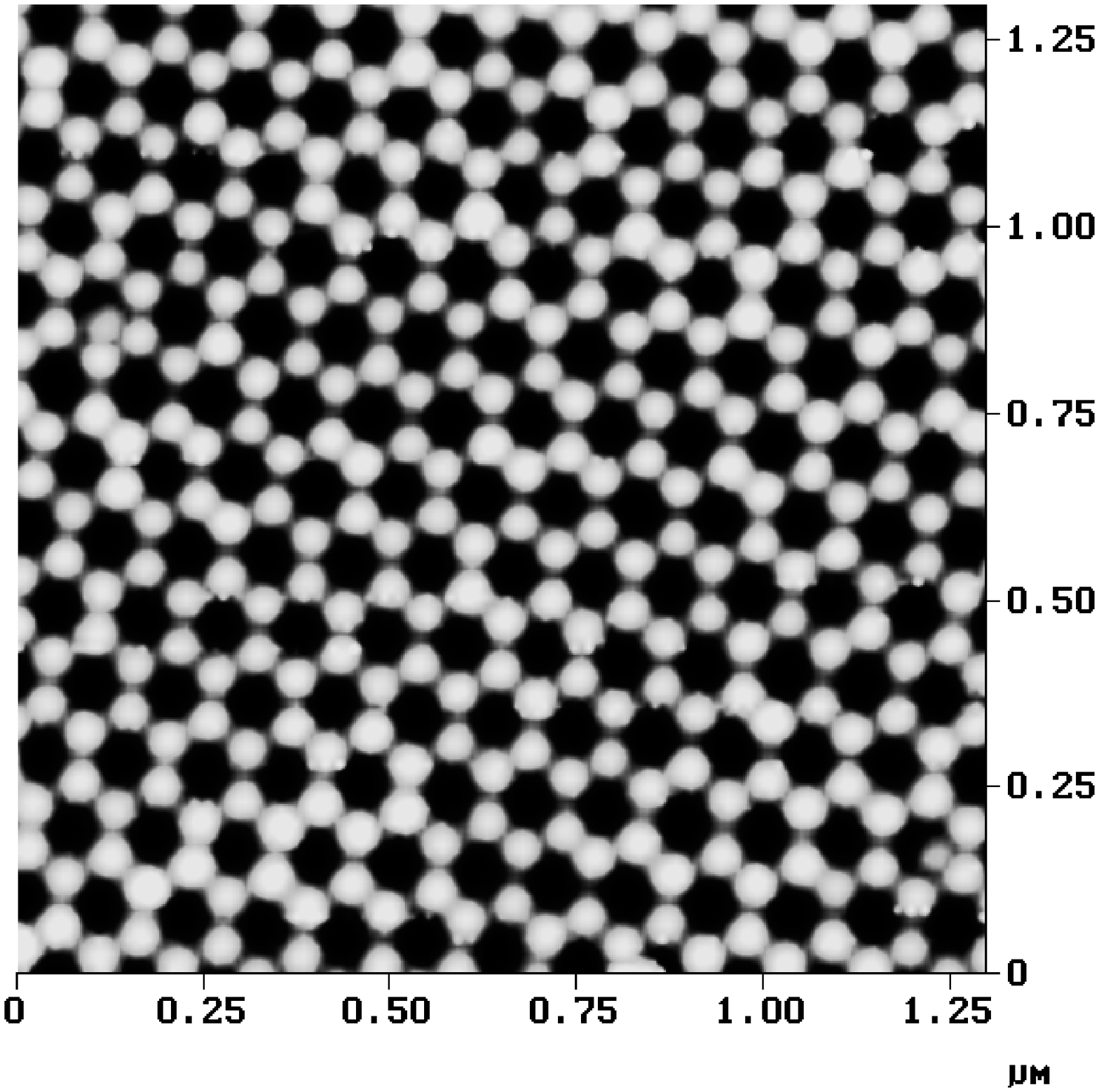,height=3.4in,width=3.4in}}
\caption{Atomic force micrograph of a porous template self assembled for 
synthesizing a 
hexagonal close packed array of quantum dots. The dark areas are pores of $\sim$ 
50 nm diameter into which a semiconductor is electrodeposited to form 50-nm 
diameter quantum dots. The surrounding white areas are an insulator. Using low 
temperature self assembly, it is possible to reduce the pore diameter to about 5 
nm or smaller, resulting in a dot density of $\sim$ 10$^{12}$/cm$^2$.}
\end{figure}
Let us now assume that we can flip the spin in a quantum dot in 1 psec. Then the 
minimum power that will be dissipated per unit area  will exceed $kTln2 \times 
10^{12}$ /(1 psec) = 3 kW/cm$^2$ (actually most of the power will be dissipated 
in the clock cycles, which we have ignored). This dissipation is at least 30 
times more than what the Pentium IV chip dissipates \cite{suman}.
Although removal of 1 kW/cm$^2$ of heat from a chip was demonstrated almost 
two decades ago, removing 3 kW/cm$^2$ from a chip is still a major challenge in 
heat sinking.

The obvious way to overcome (or, rather, circumvent) this challenge is to 
develop 
reversible logic gates that are not constrained by the Landauer $kTln2$ barrier.
In 1996, we  devised a logically and physically reversible quantum inverter 
using two exchange coupled spins \cite{icsmm}. This device is very similar to 
the single electron parametron
idea \cite{korotkov} and can be viewed as a single spin parametron. Since either 
spin could exist in a coherent superposition of two orthogonal spin states (call 
them ``upspin'' and ``downspin'' states), this would also be a ``qubit''. Later, 
Loss and DiVincenzo  devised a universal quantum gate using two exchange coupled 
spins in two closely spaced quantum dots \cite{loss}. Recently, experimental 
demonstration of coherent transfer of electron spins between quantum dots 
coupled by conjugated molecules has been demonstrated, opening up real 
possibilities in this area \cite{awschalom}. 

\section{Spintronic quantum gates}

The idea of using a single electron or nuclear spin to encode a qubit and then 
utilizing this to realize a universal quantum gate, has taken hold 
\cite{privman,kane,loss,bandy1}. The motivation for this is the realization that 
spin coherence times 
in solids is much larger than charge coherence time. Charge coherence times 
in semiconductors tend to saturate to about 1 nsec as the temperature is lowered 
\cite{mohanty}. This is presumably due to coupling to zero point motion of 
phonons which cannot be eliminated by lowering temperature \cite{mohanty}. On 
the other hand, electron spin coherence times of 100 nsec in GaAs at 5 K has 
already been reported \cite{kikkawa} and much higher coherence times are 
expected for nuclear spins in silicon \cite{feher}. Therefore, spin is obviously 
the preferred 
vehicle to encode qubits in solids. Using spin to carry out all optical quantum 
computing has also appeared as a viable and intriguing idea \cite{calarco}. The 
advantage of the all-optical scheme over the electronic scheme is that we do not 
have to read single electron spins {\it electrically} to read a qubit. 
Electrical read out is extremely difficult \cite{milburn}, although some schemes 
have been proposed for this purpose \cite{kane1,loss2,bandy3}.
Recently, some experimental progress has been made in this direction 
\cite{tarucha}, but reading a single qubit in the solid state still remains 
elusive,. The difficult part is that electrical read out requires making 
contacts to individual quantum dots, which is an engineering challenge. In 
contrast, optical read out does not require contacts. The qubit is read out 
using a quantum jump technique \cite{imamoglu} which requires monitoring the 
fluorescence from a quantum dot. Recently, it has been verified experimentally 
that the spin state of an electron in a quantum dot can be read by circularly 
polarized light \cite{cortez}. Therefore, optical read out appears to be a more 
practical approach.

\section{Conclusion}

In this article we have provided a brief history of the use of single electron 
spin in computing. We have indicated where and why spin may have an advantage 
over charge in implementing the type of devices and architectures discussed 
here.

This work is supported by the US Air Force Office of Scientific Research under 
grant FA9550-04-1-0261.

\pagebreak

\end{document}